# Phase selection and texturing in molybdenum oxide films grown by reactive magnetron sputtering

Faezeh A. F. Lahiji[1,4], Biplab Paul[1,2], Grzegorz Greczynski[1], Ganpati Ramanath[1,3,4,5], Arnaud le Febvrier[4], Per Eklund[1,4,5]

[1]*Thin Film Physics Division, Department of Physics, Chemistry and Biology, (IFM), Linköping University, SE-58183 Linköping, Sweden*
[2]*PLATIT AG, Eichholzstrasse 9, 2545 Selzach, Switzerland*
[3]*Department of Materials Science and Engineering, Rensselaer Polytechnic Institute, Troy, NY 12180, USA*
[4]*Inorganic Chemistry, Department of Chemistry – Ångström Laboratory, Uppsala University, Box 538, SE-751 21 Uppsala, Sweden*
[5]*Wallenberg Initiative Materials Science for Sustainability, Department of Chemistry – Ångström Laboratory, Uppsala University, Box 538, SE-751 21 Uppsala, Sweden*

## Abstract

Molybdenum oxide films offer a rich variety of properties for diverse applications, but exclusive synthesis of desired phases is a major challenge. Here, we demonstrate that oxygen flow ratio $f_{O_2}$ = [O$_2$]/[Ar+O$_2$] is crucial not only for phase selection of non-layered monoclinic MoO$_2$ and layered orthorhombic α-MoO$_3$ but also for controlling grain size and preferred orientation. Both mica and sapphire support exclusive MoO$_2$ formation in the $0.15 \leq f_{O_2} \leq 0.25$ window at deposition temperatures $T_{dep}$ = 400 and 500 °C, and α-MoO$_3$ formation in the $0.35 < f_{O_2} \leq 0.5$ window at 400 °C. Within $f_{O_2}$ windows favoring exclusive phase formation, high $f_{O_2}$ fosters large grains with out-of-plane *0k0* texture, except for MoO$_2$ films on c-sapphire that show no systematic trends. These findings provide a framework for rational synthesis of phase-pure monoclinic MoO$_2$ and orthorhombic MoO$_3$ with control over texture and microstructure to access desired properties.

## Introduction

Molybdenum oxide thin films are attractive for a variety of applications in electro/photo-chromic coatings [1,2], resistive memories [3,4], displays [5] and gas sensing [6]. Exclusive phase selection is crucial because optoelectronic properties [7] depend on the Mo oxidation state and $MoO_x$ stoichiometry, but is a challenge because of the rich variety of phases and polymorphs in the Mo-O system. One can obtain multiple $Mo_nO_{3n-1}$ Magnéli phases [8,9] with $4 \leq n \leq 13$ [10] besides the monoclinic $MoO_2$ [8,9,11] and orthorhombic $MoO_3$ [1,12], all of which offer vastly different properties. For instance, $MoO_2$ with $Mo^{4+}$ is a metallic conductor while $MoO_3$ with $Mo^{6+}$ is an optically transparent insulator. Molybdenum oxide films can be synthesized by many methods including wet-chemical routes [13–15], spray pyrolysis [16], and chemical [17] and physical [7,18,19] vapor deposition. Reactive magnetron sputtering of a Mo target with an oxygen/argon plasma allows exclusive $MoO_x$ phase formation by suitable choice of substrate, deposition temperature $T_{dep}$ and oxygen flow [20] to counter the tendency to form multiple phases.

This work shows the roles of oxygen flow ratio $f_{O_2} = [O_2]/[Ar+O_2]$ and deposition temperature $T_{dep}$ on the selective formation of monoclinic $MoO_2$ and orthorhombic $\alpha$-$MoO_3$. We find that $f_{O_2}$ influences not only phase selection, but also the grain orientation and microstructure. Exclusive formation of monoclinic $MoO_2$ and orthorhombic $\alpha$-$MoO_3$ are supported at specific $f_{O_2}$ ranges within $0.1 \leq f_{O_2} \leq 0.50$ on both mica and sapphire for $T_{dep} = 400$ and 500 °C. Within the exclusive phase formation windows, high $f_{O_2}$ favors large grains and strong out-of-plane *0k0* textures of monoclinic $MoO_2$ and orthorhombic $\alpha$-$MoO_3$, compared to smaller low-textured grains at low $f_{O_2}$. These insights should pave the way for the exclusive synthesis of phase-selected $MoO_x$ films with control over grain size and texture for accessing and tuning desired properties for applications.

## Experimental details

Molybdenum oxide thin films were grown using pulsed dc reactive magnetron sputter deposition from a 99.99% pure 50-mm-diameter Mo target from Plasmaterials. The Mo target was powered with bipolar pulsed-dc voltage pulses at 150 W and 100 kHz with a 2 μs reverse time and 80% duty cycle to inhibit arcing. The substrates were $10 \times 10$ mm$^2$ pieces of fluorphlogopite $KMg_3(AlSi_3O_{10})F_2(001)$ mica --referred henceforth as f-mica-- purchased from Continental Trade, and c-plane sapphire(0001) --referred henceforth as c-sapphire-- purchased from Alineason. The substrates were mounted on a rotatable sample holder in a $3\times10^{-6}$ Pa base pressure UHV sputter-deposition chamber described elsewhere [21]. Immediately prior to loading, fresh f-mica surfaces were exposed by scotch-tape exfoliation. Wet-chemical treatments were eschewed to obviate solvent intercalation and vacuum degradation. The c-sapphire substrates were ultrasonicated successively in acetone and ethanol for 5 minutes each and blow-dried with $N_2$.

Molybdenum oxide films were deposited by adjusting the oxygen flow ratio $f_{O_2}= [O_2]/[Ar+O_2]$ in the $0.05 \leq f_{O_2} \leq 0.50$ range. All depositions were carried out for 30 minutes at 2.5 mTorr pressure with total gas flow fixed at $60 \pm 1.5$ sccm. Prior to each deposition, the Mo target was sputter-cleaned at 1.7 mTorr Ar pressure at 150 W power for 2 minutes. The substrates were preheated to $T_{dep}$ = 400 °C or 500 °C and held for 15 minutes for temperature homogenization.

Symmetric θ-2θ X-ray diffraction (XRD) scans were acquired in a PANalytical X'Pert PRO diffractometer with a Cu $K_\alpha$ beam (λ = 1.54 Å) source operated at 45 kV and 40 mA. The incident optics include a 0.5˚ divergence slit, a 0.5˚ anti-scatter slit, and a Ni filter to minimize Cu $K_\beta$. The diffracted beam included a 5.0 mm anti-scatter slit and 0.04 rad Soller slits. The PreFIX stage, equipped with an X'Celerator detector, was set to acquire data during θ-2θ scans with a 0.0167°/step size and equivalent time/step of 24.76 using the PIXcel 1D detector.

Film morphology was characterized by a Leo 1550 Gemini, Zeiss scanning electron microscope (SEM) operated at 4 kV using in- and off-lens detectors to map Z-contrast and topography, respectively. X-ray Photoelectron Spectroscopy (XPS) was conducted to determine the oxidation state of Mo using a Kratos Analytical Axis Ultra DLD system with a monochromatic 1486.6 eV $AlK_\alpha$ source. The base pressure during measurements was $1.1\times10^{-9}$ Torr ($1.5\times10^{-7}$ Pa). All spectra were recorded at 150 W anode power and normal emission angle. Setting a 20 eV analyzer pass energy results in a 0.55 eV full-width- half-maximum for the Ag *$3d_{5/2}$* peak in a reference sample containing sputter-deposited Au, Ag, and Cu. The spectrometer calibration was verified by comparing Au $4f_{7/2}$, Ag $3d_{5/2}$ and Cu $2p_{3/2}$ positions to the recommended ISO standards for monochromatic Al Kα [22,23]. The spectra were acquired over a $0.3 \times 0.7$ mm$^2$ area and charge-referenced to the Fermi level $E_F$ = 0 eV [24]. We present results recorded from samples in the as-received state.

## Results and Discussion

### Monoclinic MoO₂ formation at 500 °C

Diffractograms from molybdenum oxide films deposited on f-mica (Fig.1a) with the lowest $f_{O_2}$ = 0.05 and the highest $f_{O_2}$ = 0.5 exhibit only the *00l* f-mica substrate peaks ($4 \leq l \leq 14$), suggesting amorphous MoO$_x$ formation. Multiple Bragg reflections from monoclinic MoO$_2$ [8] were detected for $0.1 \leq f_{O_2} \leq 0.35$ with relative intensities varying with $f_{O_2}$. For $f_{O_2}$ = 0.30 and 0.35, only the *020* and *040* MoO$_2$ peaks are seen, indicating that the *0k0* planes in monoclinic MoO$_2$ crystals are preferentially aligned with *00l* planes in f-mica. The lack of crystalline phase formation at high $f_{O_2}$ is likely due to oxygen-poisoning of the Mo target [25–27].

Diffractograms from films on c-sapphire (Fig.1b) show monoclinic MoO$_2$ formation for $0.05 \leq f_{O_2} \leq 0.25$. Outside this range, the exclusive presence of *00l* c-sapphire reflections suggests possible amorphous MoO$_x$ formation. For $f_{O_2}$ = 0.05, only the *00l* and *0k0* reflections from MoO$_2$ are seen, suggesting exclusive phase formation. An additional peak from hexagonal MoO$_2$ [28] detected at $f_{O_2}$ = 0.1, underscores the sensitivity of phase selection to oxygen flow. The one-dimensional phase diagram with representative interplanar spacings of MoO$_x$ phases formed plotted versus $f_{O_2}$ (see Fig. 1c-d) shows that monoclinic MoO$_2$ is exclusively favored on f-mica for $0.1 \leq f_{O_2} \leq 0.35$, and on c-sapphire for $0.05 \leq f_{O_2} \leq 0.25$. However, at $f_{O_2}$ = 0.1, distinct diffraction peaks corresponding to hexagonal MoO$_2$ are observed.

### Monoclinic MoO₂ and orthorhombic α-MoO₃ for mation at 400 °C

At $T_{dep}$ = 400 °C, the films grown on f-mica (Fig. 2a) at $f_{O_2}$ = 0.05 show no detectable MoO$_x$ reflections, suggesting possible amorphous MoO$_x$ formation, reminiscent of the behavior at $T_{dep}$ = 500 °C. The *0k0*, *0kl,* and *hk0* Bragg reflections from monoclinic MoO$_2$ are observed for $0.1 \leq f_{O_2} \leq 0.25$, indicating exclusive phase selection. For $f_{O_2}$ = 0.30, three phases are observed: monoclinic MoO$_2$ indicated by the *0k0* reflections, traces of monoclinic Mo$_8$O$_{23}$ [29] specified by the *h0l* reflection, and orthorhombic α-MoO$_3$ [30] indicated by *hkl, 0k0,* and *h0l* peaks. At $f_{O_2}$ = 0.35, only orthorhombic α-MoO$_3$ peaks are detected for $0.35 < f_{O_2} \leq 0.5$.

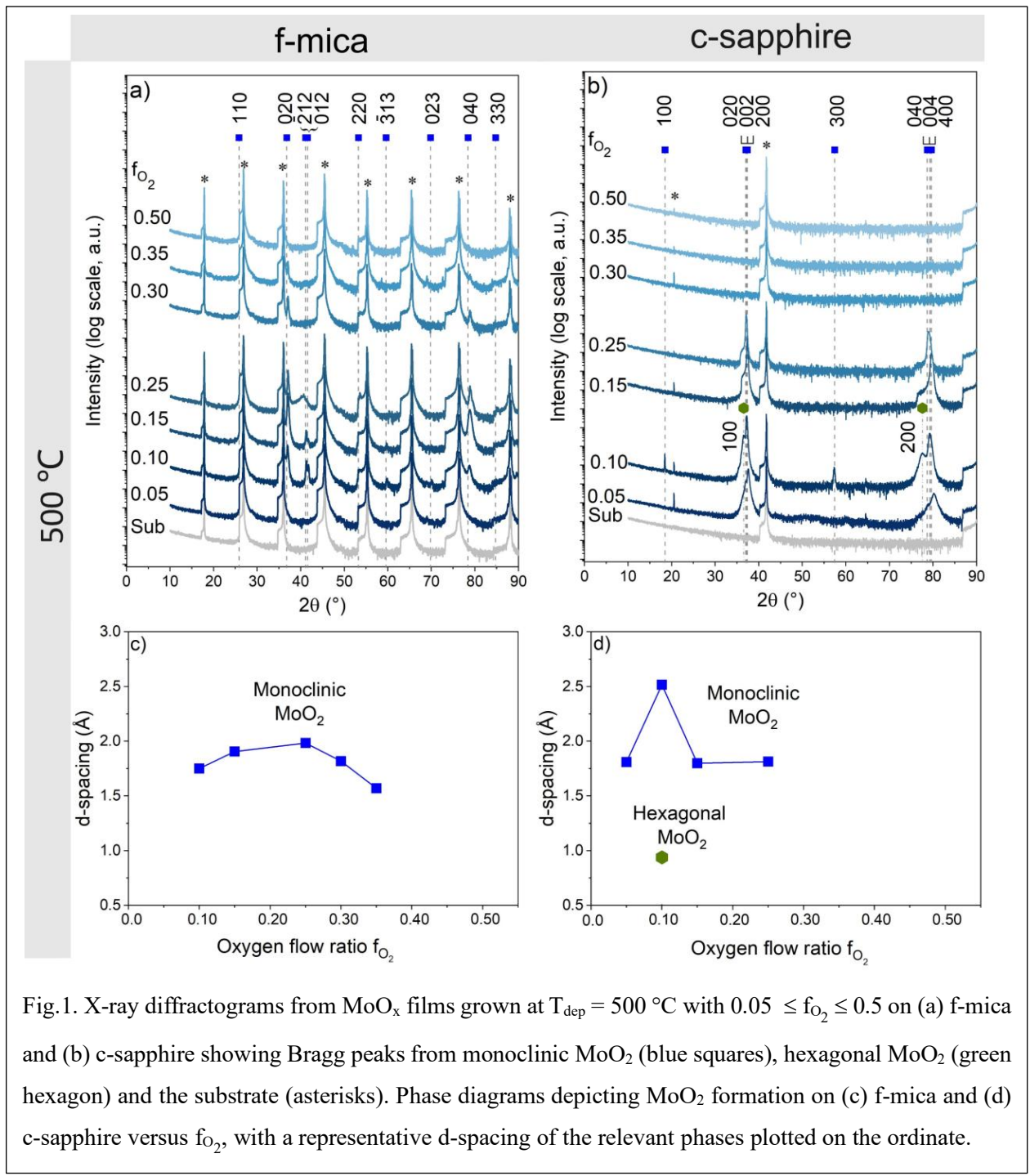

Fig.1. X-ray diffractograms from MoO$_x$ films grown at T$_{dep}$ = 500 °C with 0.05 ≤ f$_{O_2}$ ≤ 0.5 on (a) f-mica and (b) c-sapphire showing Bragg peaks from monoclinic MoO$_2$ (blue squares), hexagonal MoO$_2$ (green hexagon) and the substrate (asterisks). Phase diagrams depicting MoO$_2$ formation on (c) f-mica and (d) c-sapphire versus f$_{O_2}$, with a representative d-spacing of the relevant phases plotted on the ordinate.

Except for minor differences, MoO$_x$ films grown on c-sapphire at 400 °C and 500 °C show similar behaviors. For 0.05 ≤ f$_{O_2}$ ≤ 0.1, we predominantly observe *00l* and *h00* peaks from monoclinic MoO$_2$ (Fig. 2b) together with traces of *hk0* and *hkl* reflections from orthorhombic

Mo$_4$O$_{11}$ [31]. At f$_{O_2}$ = 0.1, we observe an additional unindexed peak at 2θ = 76.64° not associated with known MoO$_x$ phases. For 0.15 < f$_{O_2}$ ≤ 0.25, we exclusively observe monoclinic MoO$_2$.

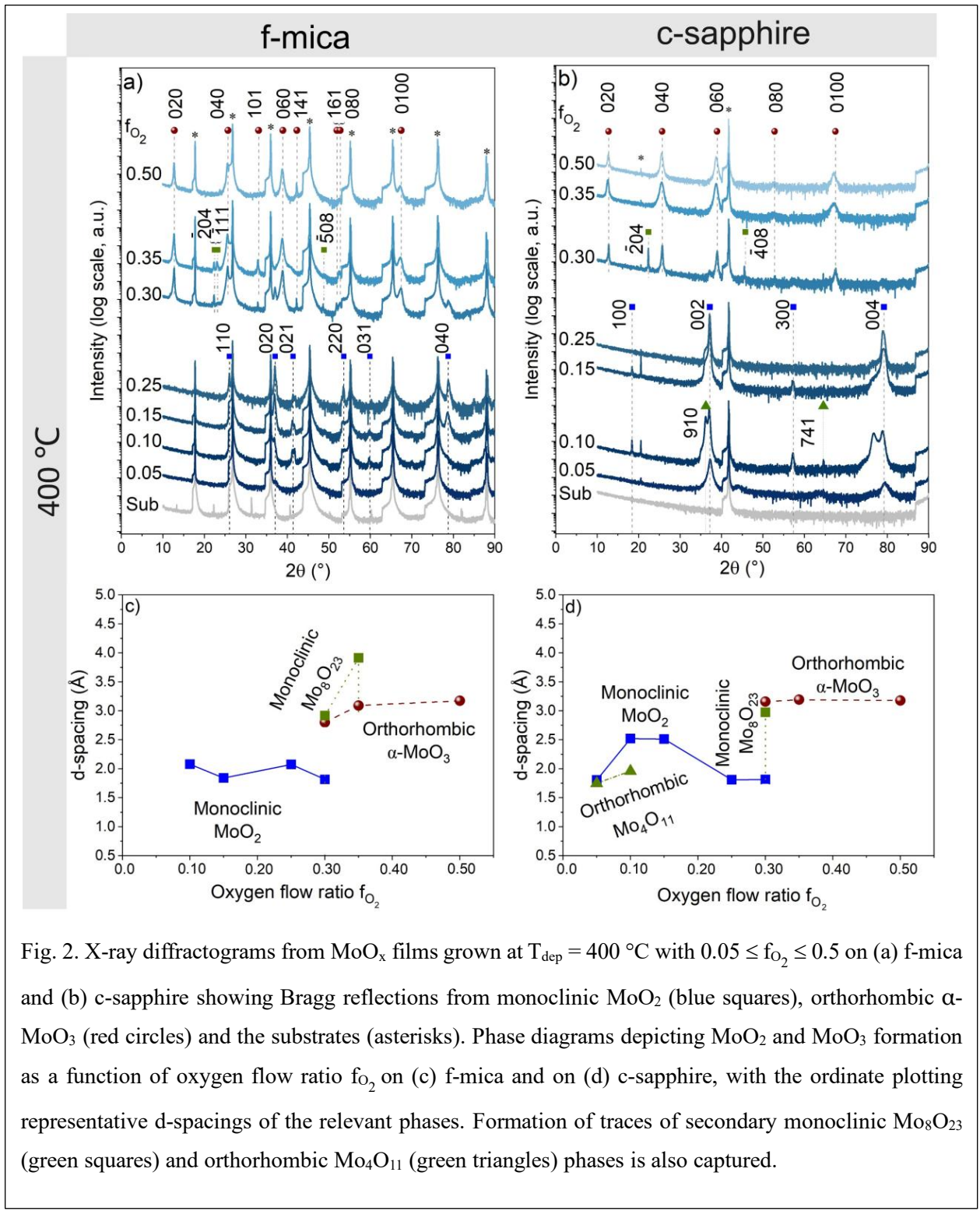

Fig. 2. X-ray diffractograms from MoO$_x$ films grown at T$_{dep}$ = 400 °C with 0.05 ≤ f$_{O_2}$ ≤ 0.5 on (a) f-mica and (b) c-sapphire showing Bragg reflections from monoclinic MoO$_2$ (blue squares), orthorhombic α-MoO$_3$ (red circles) and the substrates (asterisks). Phase diagrams depicting MoO$_2$ and MoO$_3$ formation as a function of oxygen flow ratio f$_{O_2}$ on (c) f-mica and on (d) c-sapphire, with the ordinate plotting representative d-spacings of the relevant phases. Formation of traces of secondary monoclinic Mo$_8$O$_{23}$ (green squares) and orthorhombic Mo$_4$O$_{11}$ (green triangles) phases is also captured.

At $f_{O_2}$ = 0.3, orthorhombic MoO$_3$ and monoclinic Mo$_8$O$_{23}$ are observed with traces of monoclinic MoO$_2$ similar to that seen on f-mica. Only orthorhombic MoO$_3$ forms for 0.35 ≤ $f_{O_2}$ ≤ 0.5.

XRD at 400 °C (Figs. 2c-d) indicates that MoO$_2$ is favored at low $f_{O_2}$, and α-MoO$_3$ is favored at high $f_{O_2}$, with a tendency for secondary phases at intermediate $f_{O_2}$. On f-mica, monoclinic MoO$_2$ forms exclusively at 0.1 ≤ $f_{O_2}$ ≤ 0.25 while orthorhombic α-MoO$_3$ is the sole phase observed at $f_{O_2}$ = 0.50. Both phases form at $f_{O_2}$ = 0.30, and $f_{O_2}$ = 0.35 features traces of other polymorphs. On c-sapphire, exclusive monoclinic MoO$_2$ formation is restricted to 0.15 ≤ $f_{O_2}$ ≤ 0.25, while exclusive α-MoO$_3$ formation is seen for a broader range of 0.35 ≤ $f_{O_2}$ ≤ 0.50. Traces of other secondary phases are seen for 0.05 ≤ $f_{O_2}$ ≤ 0.10 and $f_{O_2}$ = 0.30.

*Mo-O bonding chemistry oxidation state*

XPS spectra recorded from as received MoO$_x$ films grown on f-mica and c-sapphire substrates at $T_{dep}$ = 400 °C are shown in Fig. 3. Films grown at $f_{O_2}$ = 0.25 exhibit prominent Mo$^{+4}$ spin-split doublet with 3d$_{5/2}$ - 3d$_{3/2}$ peaks at 229.4 eV and 232.5, respectively on both f-mica and c-sapphire (Fig. 3a and b), consistent with monoclinic MoO$_2$ indicated by XRD results. Lower-intensity 3d$_{5/2}$ - 3d$_{3/2}$ doublets at 231.0 / 234.2 eV and 232.5 /235.65 eV correspond to Mo$^{+5}$ and Mo$^{+6}$ oxidation states, respectively [32–34]. Since XRD results show only MoO$_2$ Bragg reflections at $f_{O_2}$ = 0.25, these higher oxidation states may occur in the amorphous or nanocrystalline phase and/or be restricted to the surface.

Films grown on f-mica with $f_{O_2}$ = 0.35 reveal nearly the exclusive presence of Mo$^{+6}$ state (Fig. 3c-d) on f-mica and c-sapphire, indicating the MoO$_3$ phase, consistent with the XRD results. In particular, the Mo 3d$_{5/2}$/3d$_{3/2}$ doublet at 232.9 / 236.0 eV which is associated with a bluish-white coloration agrees with the reported value for MoO$_3$ [35,36]. Additionally, the presence of Mo$^{+5}$ oxidation state, indicated by peaks located at 231.5/234.6 ± 0.05 eV for 3d$_{5/2}$ and 3d$_{3/2}$, respectively is associated with oxygen vacancies and positively charged structural defects in MoO$_3$ lattice on both substrates, with a more pronounced intensity on c-sapphire [35,37] .

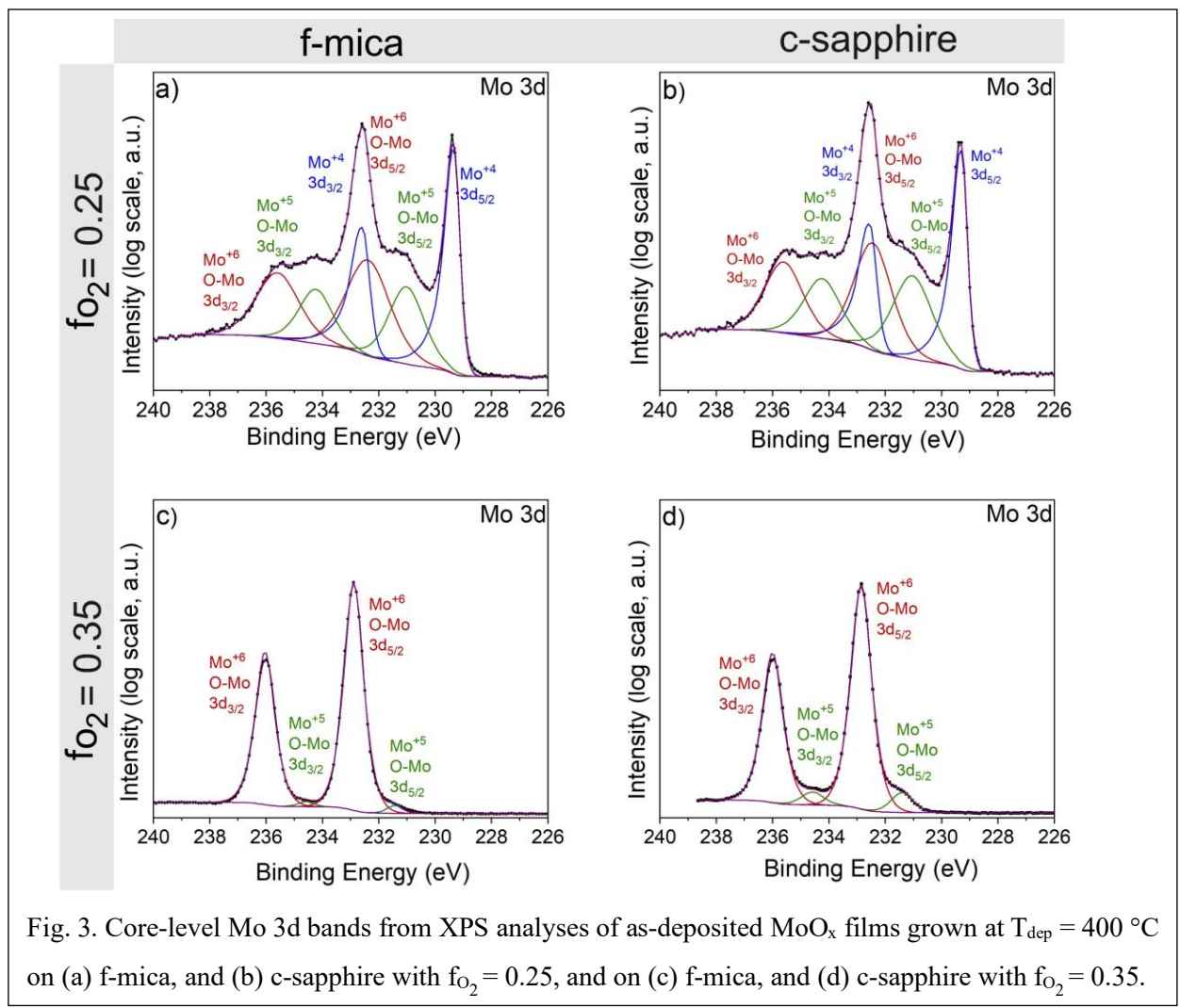

Fig. 3. Core-level Mo 3d bands from XPS analyses of as-deposited MoO$_x$ films grown at T$_{dep}$ = 400 °C on (a) f-mica, and (b) c-sapphire with f$_{O_2}$ = 0.25, and on (c) f-mica, and (d) c-sapphire with f$_{O_2}$ = 0.35.

### *Preferred orientation as a function of f$_{O_2}$*

Monoclinic MoO$_2$ films deposited on f-mica at 500 °C show increasing out-of-plane *020* texture with increasing f$_{O_2}$ (Fig. 4a). The non-*0k0* MoO$_2$ reflections of comparable intensity seen at low f$_{O_2}$ suggest diverse grain orientations. As f$_{O_2}$ increases, the non-*0k0* reflections decrease in number and completely disappear at f$_{O_2}$ = 0.30 even as the *020* MoO$_2$ reflection intensities continue to increase with f$_{O_2}$ for 0.30 ≤ f$_{O_2}$ ≤ 0.35. The trends are similar for T$_{dep}$ = 400 °C (Fig. 4b) but are somewhat obscured by the presence of additional phases. The f$_{O_2}$-driven accentuation of out-of-plane *020* texture indicates the tendency of *020* planes in the MoO$_2$ crystals align with f-mica *00l*.

MoO$_2$ films on c-sapphire show no T$_{dep}$ - f$_{O_2}$ windows with texture-f$_{O_2}$ correlations. Films grown at T$_{dep}$=500 °C (Fig. 4c) feature dominant *002* MoO$_2$ reflections for f$_{O_2}$= 0.05, and f$_{O_2}$ =0.1, with smaller *020* MoO$_2$. Higher f$_{O_2}$ =0.15, and f$_{O_2}$ =0.25, result in solitary *200* and *002* peaks, respectively. Films grown at T$_{dep}$= 400 °C (Fig. 4d) with f$_{O_2}$ = 0.05 and 0.25 showing solitary *002* and *020* peaks. The *002* peak dominates over smaller *100* reflection for f$_{O_2}$ =0.1, and 0.15.

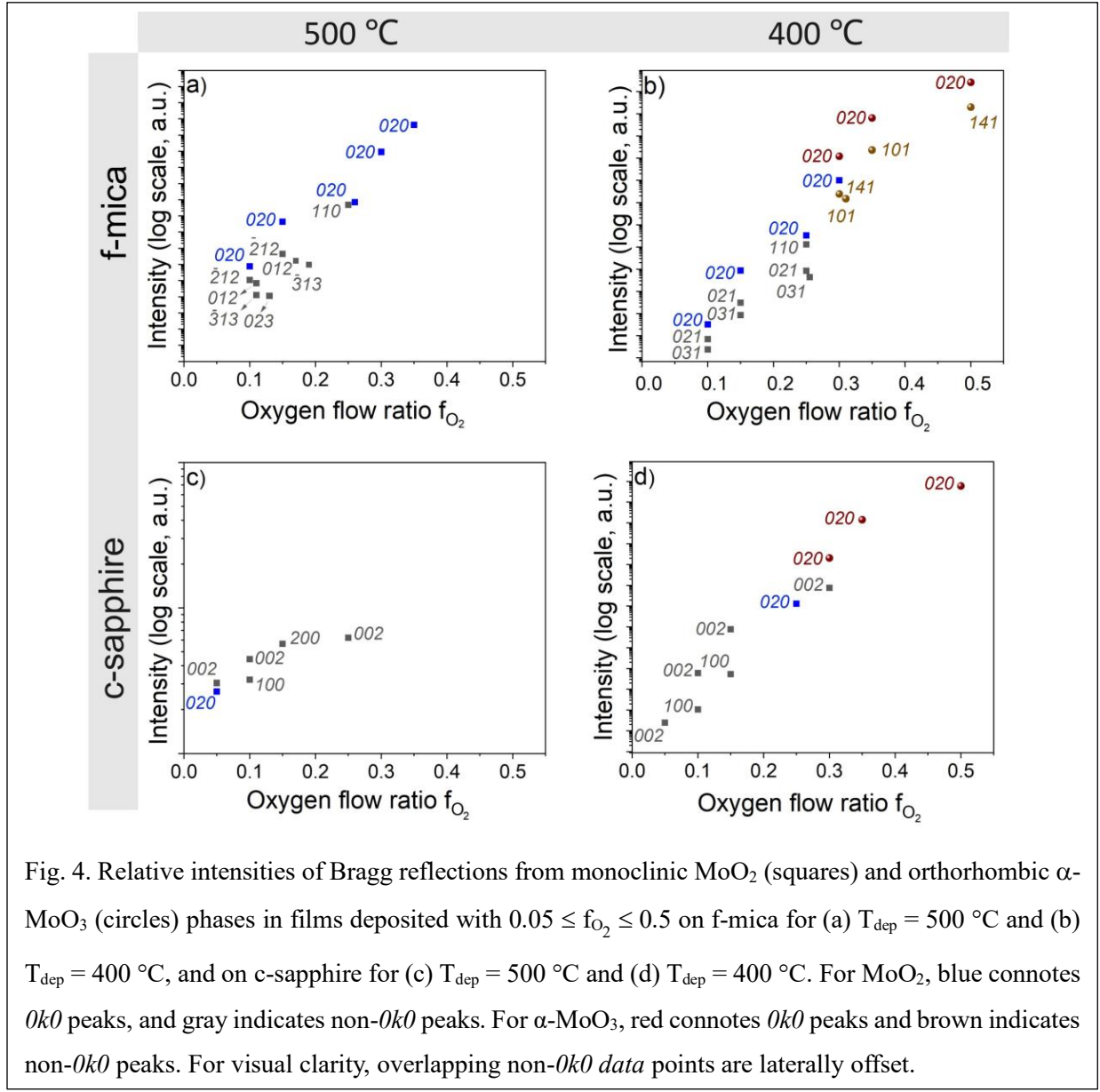

Fig. 4. Relative intensities of Bragg reflections from monoclinic MoO$_2$ (squares) and orthorhombic α-MoO$_3$ (circles) phases in films deposited with 0.05 ≤ f$_{O_2}$ ≤ 0.5 on f-mica for (a) T$_{dep}$ = 500 °C and (b) T$_{dep}$ = 400 °C, and on c-sapphire for (c) T$_{dep}$ = 500 °C and (d) T$_{dep}$ = 400 °C. For MoO$_2$, blue connotes *0k0* peaks, and gray indicates non-*0k0* peaks. For α-MoO$_3$, red connotes *0k0* peaks and brown indicates non-*0k0* peaks. For visual clarity, overlapping non-*0k0 data* points are laterally offset.

Orthorhombic α-MoO$_3$ films deposited with 0.30 ≤ f$_{O_2}$ ≤ 0.50 at 400 °C show a strong out-of-plane *0k0* texture on both f-mica (Fig. 4b) and c-sapphire (Fig. 4d). Out-of-plane *020* texture indicates the tendency of the b-axis of the α-MoO$_3$ crystals to orient along the surface normal of both f-mica and c-sapphire substrates. The monotonic increase in the *0k0* α-MoO₃ intensity for films on c-sapphire for $0.30 \leq f_{O_2} \leq 0.5$ indicates enhanced *0k0* texturing with increasing $f_{O_2}$.

The above results collectively indicate that high $f_{O_2}$ correlates with higher *0k0* textures in MoO₂ on f-mica and MoO₃ films on both f-mica and c-sapphire for $f_{O_2}$-$T_{dep}$ windows of exclusive formation of these phases. These texture-$f_{O_2}$ correlations seen for these phase-substrate combinations are valuable for tailoring texture by adjusting $f_{O_2}$.

*Microstructures of MoO₂ and MoO₃ films*

Films exclusively containing either MoO₂ or MoO₃ phases exhibit distinctive microstructures on f-mica and c-sapphire. While the MoO₂ microstructure was sensitive to both the substrate and $T_{dep}$, nearly identical MoO₃ microstructures were obtained on both substrates. Monoclinic MoO₂ films on f-mica at $T_{dep}$ = 500 °C and $f_{O_2}$ = 0.25 reveal ~190-nm-sized plate-shaped grains with ~25-nm-thick plate edges-oriented outward from the surface plane (Fig. 5a). MoO₂ films on c-sapphire (Fig. 5b) show coarser anisotropic ~34-nm-wide ~130-nm-long prism-shaped crystals merging with each other, similar to that reported sputter-deposited MoO₂ films [38]. At $T_{dep}$ = 400 °C for the same $f_{O_2}$ = 0.25, MoO₂ films on both f-mica and c-sapphire exhibit similar microstructures (Figs. 5c-d) with finer grain sizes than seen at $T_{dep}$ = 500 °C. In particular, the plate-shaped grains of MoO₂ on f-mica are ~16-nm-wide and ~57-nm-thick. The prism-shaped MoO₂ crystals c-sapphire are ~24-nm-wide and ~70-nm-long. The finer grain structure at lower $T_{dep}$ is consistent with lower surface adatom mobilities and diffusion distances during film growth. Orthorhombic α-MoO$_3$ films obtained at $f_{O_2}$ = 0.35 consist of large 200-600-nm-wide sheet-shaped grains on both substrates (Figs. 5e-f).

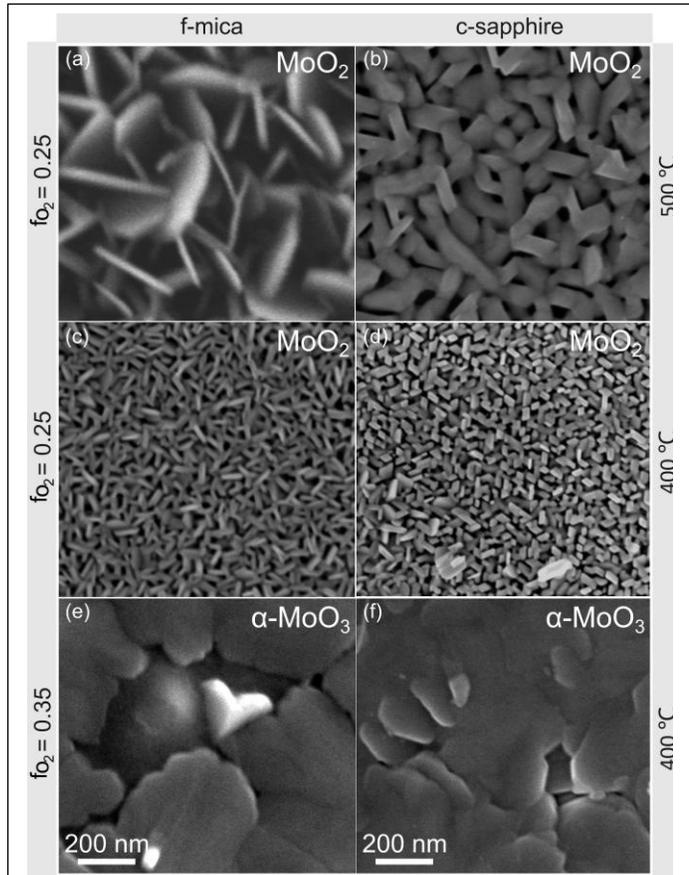

Fig. 5. SEM micrographs from films with either exclusive monoclinic MoO₂ or orthorhombic MoO₃ phases obtained at specific $f_{O_2}$-$T_{dep}$ combinations on f-mica and c-sapphire.

These microstructures of MoO$_2$ and MoO$_3$ featuring distinctively shaped crystals point to the connection between crystal shape and preferred orientation. For instance, the out-of-plane *0k0* MoO$_2$ texture on f-mica indicates a preference for *0k0* planes to stack along the thinnest dimension of the plate-shaped grains. The prism-shaped grains in MoO$_2$ films on c-sapphire correlate with the presence of prominent *h00* and *00l* peaks. The large flat sheets of α-MoO$_3$ are consistent with a strong out-of-plane *0k0* texture on both substrates.

## Conclusions

Oxygen flow ratio $f_{O_2}$ is a key factor in determining phase selection, texture, and microstructure of MoO$_x$ films grown on f-mica and c-sapphire by reactive magnetron sputtering. Phase-pure monoclinic MoO$_2$ and orthorhombic α-MoO$_3$ films can be exclusively obtained by adjusting $f_{O_2}$ and the deposition temperature $T_{dep}$. Broadly, low $f_{O_2}$ favors MoO$_2$ formation at $T_{dep}$ ~400 and 500 °C, while high $f_{O_2}$ favors α-MoO$_3$ formation at the lower temperature. Within the $f_{O_2}$ windows of exclusive formation of MoO$_2$ and α-MoO$_3$ films, higher $f_{O_2}$ fosters *0k0* texture, except for MoO$_2$ films on c-sapphire that show no systematic trends. These findings provide a framework for synthesis of phase-pure monoclinic MoO$_2$ and orthorhombic MoO$_3$ with control over texture and microstructure.

## Acknowledgements

The authors acknowledge funding from the Swedish Government Strategic Research Area in Materials Science on Functional Materials at Linköping University (Faculty Grant SFO-Mat-LiU No. 2009 00971), the Knut and Alice Wallenberg foundation through the Wallenberg Academy Fellows program (KAW-2020.0196), the Swedish Research Council (VR) under Project No. 2021-03826, and the Swedish Energy Agency under project number 52740-1. This work was partially supported by the Wallenberg Initiative Materials Science for Sustainability (WISE) funded by the Knut and Alice Wallenberg Foundation, and the US National Science Foundation grant CMMI 2135725 through the BRITE program.